\documentclass{ws-procs9x6}

\begin{document}

\title{Pentaquark Baryons in String Theory \\-Talk at Pentaquark 04-}

\author{M. Bando}

\address{Physics Division, Aichi University, Aichi 470-0296, Japan}

\author{T. Kugo}

\address{Yukawa Institute, Kyoto University, Kyoto 606-8502, Japan}  

\author{A. Sugamoto and S. Terunuma}

\address{Department of Physics and Graduate School of Humanities and Sciences,\\ Ochanomizu University, Tokyo 112-8610, Japan}  

\maketitle

\abstracts{Pentaquark baryons $\Theta^{+}$ and $\Xi^{--}$ are 
studied in the dual gravity theory of QCD in which simple mass formulae of pentaquarks are derived in the Maldacena prototype model for  supersymmetric QCD and a more realistic model for  ordinary QCD.   In this approach it is possible to explain the extremely narrow decay widths of pentaquarks.  With the aim of constructing more reliable mass formulae, a sketch is given on spin and the hyperfine interaction in the string picture.}
\section{Introduction}
  We are very happy to give a talk at this pentaquark 04 conference, on the bassis of our recent papers~\cite{BKST}.

It made a great impact on us when the pentaquark baryon $\Theta^{+}$ was found here at SPring-8 by T. Nakano {\it et al.} last year.  $\Theta^{+}$ is an exotic baryon consisting of five quarks, $(ud)(ud)\bar{s}$,  having the mass, $M(\Theta^{+})=1,540 \pm 
10 \mbox{MeV}$, and the width, $\Gamma(\Theta^{+}) \leq 25 \mbox{MeV}$.  Subsequently, 
other pentaquarks, $\Xi^{--}((ds)(ds)\bar{u})$ and $\Theta^{0}_{c}((ud)(ud)\bar{c})$, were reported to be observed at CERN NA49 and HERA H1, respectively.  At this conference, we have also learned that there are positive and negative indications on the observation of these pentaquarks, depending on the experimental apparatuses.

  Pentaquarks were predicted by
Diakonov {\it et al.} in 1997 as chiral solitons.  As is well known, in the naive quark model, hadron masses are estimated as the sum of masses of constituent quarks and energy of the hyperfine or the color magnetic interactions.
   Masses of triquarks calculated in this way are in good agreement with the observed values, but for pentaquarks, the observed masses are about 200 MeV lower than the predicted values, and the observed widths are very narrow, being about 1/100 of their Q-values~\cite{Oka}.

Therefore, this is a very interesting problem to inquire.

Jaffe and Wilczek treated the penatquarks as being composed of two diquark pairs $(ud)(ud)$ and one anti-quark $\bar s$, while Karliner and Lipkin considered they are made of two clusters, diquark $(ud)$ and triquark $(ud\bar s)$. The pentaquarks were also studied in lattice QCD and QCD sum rules.  In this conference, we had a number of good talks on the QCD flux tube models of pentaquarks, which were presented by Y. Enyo, E. Hiyama, S. Takeuchi, F. Okiharu, and H. Suganuma~\cite{QCDfluxtube} 
\cite{Suganuma}.

The purpose of this talk is to study the pentaquark baryons in colored string theory, using the recent development in string theories started by J. Maldacena in 1998.  

The best way to understand this picture is to draw a picture of 
$\Theta^{+}$ as quarks connected by colored strings of red, green and blue, which gives a very beautiful shape displayed in Fig.1.

\begin{figure}
\begin{center}
\includegraphics[width=10cm,clip]{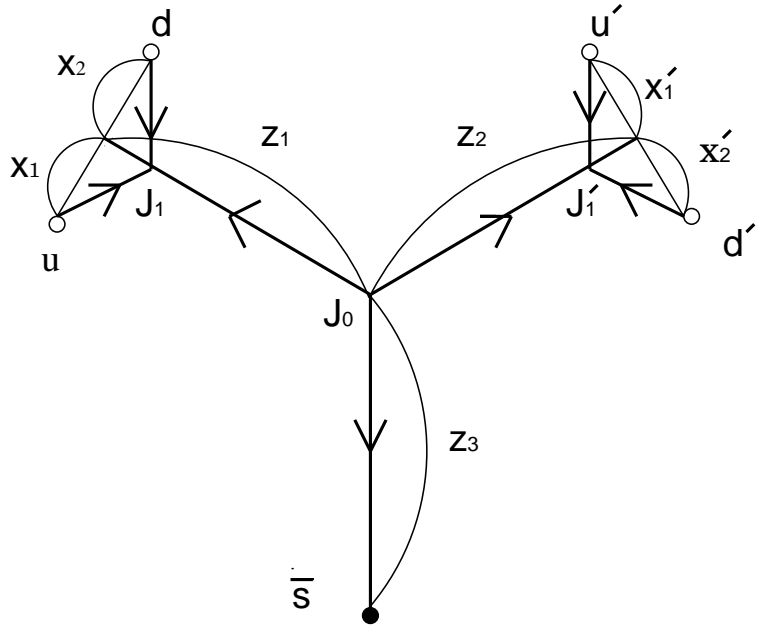}
\end{center}
\caption{Three-dimensional view of the pentaquark $\Theta^+$ in the string picture.}
\end{figure}%

In this picture, the mass of $\Theta^{+}$ is estimated as the total length of the colored strings located in the curved space with extra dimensions.  Furthermore, there may be an unexpected merit of this picture; that is, this branched web is quite stable.  This is because, for a pentaquark to decay into a meson and a triquark baryon, string configurations with a loop must appear in the intermediate stage of decay, but this may be a rare occurrence.  Intuitively speaking, in order for the pentaquark to decay, the string connecting two junctions $J_{0}$ and $J_{1(1')}$, say with red color, should be replaced by two strings with anti-blue and anti-green colors.  This replacement is energetically and stochastically very dificult to ocuur. 

\section{Dual gravity model of QCD} 
The correspondence betwen  dual gravity model and QCD, usually called AdS/CFT correspondence, is a very difficult concept for us, but it may be understood by using factorization and vacuum insertion.

QCD consists of quarks and gluons; quarks possess both color $(r, g, b)$ and flavor $(u, d, s, {\it etc.})$, while gluons possess color $(r, g, b)$ and anti-color $(\bar{r}, \bar{g}, \bar{b})$ but not flavor.  An open string (a string with two endpoints) is ideally suited to account for such quantum numbers at its two ends.  For quarks, one end represents color and the other end flavor.  For gluons, one end represents color and the other anti-color.  

In recently developed string theory, we prepare ``branes" (higher dimensional extended objects that are generalized membranes) on which the endpoints of these open strings are confined to move. Applying this idea to QCD, we introduce $N_{c} (=3)$ ``colored branes" and $N_{f}$ ``flavored branes" at which open strings corresponding to quarks and gluons terminate.  

Because the classical energy of a string is proportional to its length and because gluons are massless, $N_c$ colored branes should lie on top of one another. On the other hand, quarks possess intrinsic masses, and therefore the endpoints of a quark string, namely, a flavored brane and a colored brane should be separated from each other by a nonvanishing distance $U$.  If the direction of the separation is chosen along the fifth (extra) dimension $u$, the energy stored by this separation is the internal one.  Then, the intrinsic quark mass $m_q$ can be represented as $m_q=U\times 
(\mbox{string tension})$, where the string tension is the energy stored inside a unit length of string. 

To evaluate the amplitude for a certain process to occur in the above picture, we have to sum up all the possible two-dimensional world sheets swept by the string with the weight $\exp(iS)$, where the action $S$ is given by $S$=(energy)$\times$(time)=(area of the string's world sheet)$\times$(tension).  As stated above, the endpoints of the strings are confined to the colored branes or flavored branes, so that the world sheet has boundary trajectories $\{C_{i}\} (i=1, 2, ...)$ on colored branes and $\{F_{j}\} (j=1, 2, ...)$ on flavored branes.   This amplitude is denoted as $A(\{C_{i}\}, \{F_{j}\})$.

Let us make a simple approximation using factorization and a vacuum insertion, which is frequently used in ordinary QCD. 
For example, in the decay of $B^{0}\rightarrow K^{-}e^{+}\nu$, we factorize the current-current interaction and use the vacuum insertion $|vac\rangle\langle vac|$:  
\begin{equation}
\langle B^{0}|J^{+\mu}J^{-}_{\mu}|K^{-}, e^{+}, \nu\rangle\approx\langle B^{0}, K^{+} |J^{+\mu}| vac\rangle\langle vac|J^{-}_{\mu}|e^{+}, \nu\rangle.
\end{equation}
In the same way, the string amplitude can be approximated by the factorized amplitude with a vacuum insertion:
\begin{equation}
A(\{C_{i}\}, \{F_{j}\})\approx \langle\{C_{i}\}|vac\rangle\langle vac|\{F_{j}\}\rangle.
\end{equation}
Summing up all the possible configuration of $\{C_{i}\} (i=1, 2, ...)$ gives 
\begin{eqnarray}
A(\{F_{j}\}) \approx \left( \sum_{\{C_{i}\}}\langle\{C_{i}\}|vac\rangle \right)\times \left(\langle vac|\{F_{j}\}\rangle\right) 
\propto \langle vac|\{F_{j}\}\rangle.
\end{eqnarray}

Now, the remaining problem is to determine what the vacuum state is.  As seen from the first factor, $\sum_{\{C_{i}\}} \langle\{C_{i}\}|vac\rangle$, the existence of $N_{c}$ colored branes deform the flat vacuum to the curved space with compactification.  The various curved spaces with compactification (vacua) are known after Maldacena's work.  If we prepare  $N_c(=3)$  four dimensional Minkowski spaces (world sheets of D3-branes) for the colored branes,  the vacuum becomes the five-dimensional Anti de Sitter space $AdS_{5} \times S^{5}$. We call this Maldacena's prototype model which corresponds to $N=4$ supersymmetric $SU(N_{c})$ Yang-Mills theory, but it is not the ordinary QCD.

To describe the five-dimensional space, we introduce an extra coordinate $u$ which measures the intrinsic quark masses in addition to the Minkowski space, $(t, z, {\bf x}_{\perp})$, along which the world volume of the stuck $N_{c}$ colored branes extend.  

In more realistic models, we need to break the supersymmetries.  For this purpose, an effective method is to compactify one space-like dimension to a circle, a variant of the method of Witten.  Then, we obtain AdS Schwartzshild spaces.  In this way, the difference in boundary conditions between fermion and boson in the compactified dimension breaks the supersymmetries completely.  Therefore, if the radius of the compactified circle is $R_{KK}$, then the mass scale $M_{KK}=2\pi/R_{KK}$ is introduced.

The metric of the vacuum deformed by the existence of $N_{c}$ colored branes whose world sheets are Minkowski space times the circle, is known.  We have used this curved space to describe a more realistic model of QCD.

 \section{General formulation of Pentaquarks}
As discussed in the previous section, to evaluate the amplitude or the energy of pentaquarks, we have to evaluate 
$\langle vac|\{F_{j}\}\rangle$.  We are interested in evaluating the static energy of pentaquarks.  So, we first fix the position of flavors, or fix the static trajectories of five quarks,  $F_{j=u,d,u',d',{\bar s}}$, on the flavored branes. 
The $u$- and $d$-quarks are placed on the same flavored brane, since $u$ and $d$ have an almost equal mass.  On the other hand, $s$-quark is heavior than $u$ and $d$, so that $s$-quark is placed on another brane located farther from the colored branes than the brane of $u$ and $d$.

These five quarks are connected by colored strings as in Fig.1.  This picture shows the three dimensional view. In our treatment, however, the pentaquark is located in the five dimensional curved space determined by the dual gravity theory of QCD.  Therefore, the strings can extend also in the fifth dimension (u-direction).  This is the same problem of finding the shape and length of a string placed under the gravity, where both ends of string are picked up by hands.  In our problem the virtical coordinate corresponds to $u$, while the horizontal coordinates on the earth correspond to $x'$s and $z'$s in Fig. 1.

Therefore, we can solve this problem easily and obtain the energy stored inside the strings of pentaquarks as the function of coordinates $z'$s and $x'$s.  Subtracting the rest masses of quarks we obtain the potential of the pentaquark.

\section{Maldacena prototype model}
By solving the non-relativistic Sch\"odinger equations in the Maldacena prototype model, following the method just mentioned, the mass formula of the pentaquark family of $\Theta^{+}$ is obtained as
 \begin{equation}
M((q_1q_2)(q'_1q'_2)\bar{q}_3)=2(m_1+m_2)(A+B)+m_3 A,
\end{equation}
while that of the triquark family of nucleons reads
\begin{equation}
M(q_1q_2q_3)=(m_1+m_2+m_3)A,
\end{equation}
where $A=1-\alpha_c N_c a^2/\pi$ and $B=-\alpha_c N_c b^2/\pi$, with
$a \approx0.359$  and $b \approx 0.236$.

\section{Pentaquarks in a QCD like model}
In this model, we obtain the following mass formulae:
\begin{eqnarray}
\bar{M}(\mbox{pentaquark}) &=&\bar{m}_1+\bar{m}_2+\bar{m}_3 + 6^{\frac{2}{3}} \{(\bar{m}_1)^{\frac{2}{3}}+(\bar{m}_2)^{\frac{2}{3}} \} \nonumber\\
&+&\frac{2}{3}(\alpha_{c} N_{c})^2 \{ 2 (\bar{m}_1+\bar{m}_2)^{-\frac{1}{3}} +(\bar{m}_3)^{-\frac{1}{3}} \}, \\
\bar{M}(\mbox{triquark}) 
&=&\bar{m}_1+\bar{m}_2+\bar{m}_3 
+\frac{2}{3}(\alpha_{c} N_{c})^2 \sum_{i=1-3} (\bar{m}_{i})^{-\frac{1}{3}} .
\end{eqnarray}
Here, the pentaquark and triquark are considered to be $((q_1 q_2)^2 \bar{q}_3)$ $(q_1 q_2 q_3)$, respectively, and the dimensionaless masses with bars are normalized by $M_{KK}/(\alpha_{c}N_{c})$.

We choose the input parameters, $M(N)=$939 MeV and $M(\Sigma)=$1,193 MeV, and $M(\Sigma_{c})=$2,452MeV.  Then, fixing  $\alpha_c$ to 0.33 (or $N_c \alpha_c=1$), the quark masses and the pentaquark masses are estimated respectively as $m_u=m_d=$313--312 MeV, $m_s=$567--566 MeV, $m_c=$1,826 MeV, and  $M(\Theta)=$1,577--1,715 MeV, $M(\Xi)=$1,670--1,841 MeV, 
$M(\Theta_{c})=$2,836--2,974 MeV, $M(\Xi_{c})=$3,266--3,556 MeV,
corresponding to the KK mass scale of $M=M_{KK}=$2--5 MeV.
Here, the pentaquark masses with $c$-quark are the new predictions, not included previously~\cite{BKST}.

\section{Decay process of pentaquarks}

In the string picture, the decay processes of $\Theta^{+} \to n+K^{+}$ are displayed in Fig.2.  
The main step is the recombination of two strings.
\begin{figure}
\begin{center}
\includegraphics[width=10cm,clip]{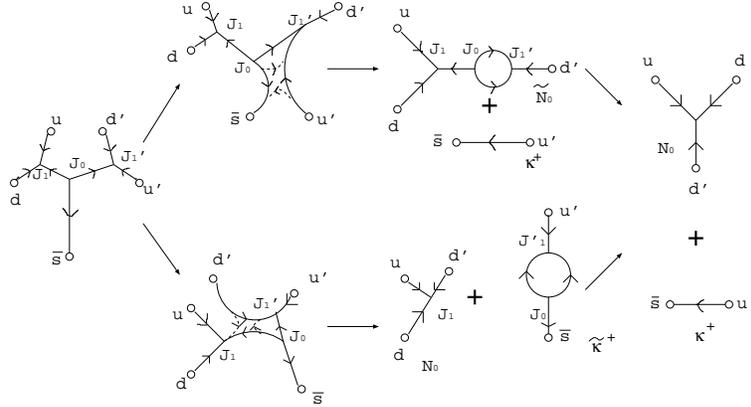}
\end{center}
\caption{Decay processes of $\Theta^+ \to K^{+} + N^0$ (neutron).}
\end{figure}%

In each channel, neutron  $n$  or a  $K$ meson accompanied by a ``string loop" is created.  This is the key point of having narrow widths for pentaquarks.

The recombination of strings can be replaced by the other process in which a string segment is firstly splitted by the pair production of quarks, producing a baryon with five quarks or a meson with four quarks.  The importance of these states are also pointed out by D. Diakonov in this conference.  Subsequently, these exotic baryon or meson becomes neutron or K meson with the string loop by the pair annihilation of quarks. 

If the state with a ``sring loop" is denoted with tilde, two decay channels can be written as follows:
\begin{equation}
\Theta^{+} \rightarrow\left\{
\begin{array}{@{\,}ll}
& \tilde{N}^{0}+K^{+}\to N^0+K^{+}~~ \mbox{(channel 1)}, \\
& N^0+\tilde{K}^{+} \to N^0+K^{+}~~ \mbox{(channel 2)}.
\end{array}
\right.
\end{equation}

The narrow width, $O(1)$ MeV, of the pentaquarks, comes from the difficulty of forming the ``string loop" states in the decay process.
Using PCAC we can show that the mass mixing between states with and whithout the string loop should be small, being roughly $1/10$ as large as their masses.

In this conference H. Suganuma~\cite{Suganuma}, starting from our decay mechanism, identified the states with a string loop to be the first gluonic excitation of hadrons.  He claimed that the excited state is  about 1 GeV heavier than the ordinary hadrons by the lattice calculation, and that the decay amplitude has the suppression factor of about 1/150.  This Suganuma's talk reinforced our suppression mechanism of pentaquark's decay.    

\section{Preliminary sketch of spin and the hyperfine interaction}

In string theory we have the fermionic variables $\psi^{\mu}(\tau, \sigma)$ in addition to the bosonic variables $X^{\mu}(\tau, \sigma)$.  The former is the distribution function of $\gamma$ matrices, so that the  ``spin" is distributed on the whole string in the string picture.  This is probably useful to explain the spin crisis of hadrons.
In this string picture we have obtained a formula of the hyperfine interactions.  Detailed anaysis of the stringy hyperfine interaction will make  the study of pentaquarks more realistic.

\section{Conclusion} 
1) Pentaquark baryons are studied in the dual gravity model of QCD.  2) This model may be understood by using factorization and vacuum insertion.  3) Simple connection conditions are derived at junctions of string webs.  4) In the extremely naive approximation, mass formula is obtained, and the decay rate is roughly estimated.  5) Nevertheless, the predictions do not differ significantly from the experiment values.  6) Spins and the hyperfine interaction are sketched in string theory in order to approach more realistic study.  7) The concept that colors and flavors are located on the ends points of strings while spins are distributed on the whole string may give new insights on hadron physics. 

Now, the string theory comes down to the real world ?!


\section*{Acknowledgments}
The authors give their sincere thanks to Prof. H. Toki, Prof. A. Hosaka, and all the staff of Pentaquark 04  for their excellent organization and for giving the opportunity of presenting this talk.

\end{document}